# Strain-induced ferromagnetism in antiferromagnetic LuMnO$_3$ thin films

J.S. White,[1,2] M. Bator,[1] Y. Hu,[1] H. Luetkens,[1] J. Stahn,[1] S. Capelli,[3] S. Das,[4] M. Döbeli,[5] Th. Lippert,[1] V.K. Malik,[4] J. Martynczuk,[6] A. Wokaun,[1] M. Kenzelmann,[1] Ch. Niedermayer,[1] and C.W. Schneider[1,*]

[1] *Paul Scherrer Institut, CH-5232 Villigen, Switzerland*

[2] *Laboratory for Quantum Magnetism, Ecole Polytechnique Fédérale de Lausanne (EPFL), CH-1015 Lausanne, Switzerland*

[3] *Institut Laue-Langevin, 6 rue Jules Horowitz, 38042 Grenoble, France*

[4] *University of Fribourg, Department of Physics and Fribourg Centre for Nanomaterials, Chemin du Musée 3, CH-1700 Fribourg, Switzerland*

[5] *Labor für Ionenstrahlphysik, Eidgenössische Technische Hochschule Zürich, CH-8093 Zürich, Switzerland*

[6] *Electron Microscopy Center of ETH Zurich (EMEZ), Eidgenössische Technische Hochschule Zürich, CH-8093 Zürich, Switzerland*

Single phase and strained LuMnO$_3$ thin films are discovered to display co-existing ferromagnetic and antiferromagnetic orders. A large moment ferromagnetism ($\approx$1$\mu$B), which is absent in bulk samples, is shown to display a magnetic moment distribution that is peaked at the highly-strained substrate-film interface. We further show that the strain-induced ferromagnetism and the antiferromagnetic order are coupled via an exchange field, therefore demonstrating strained rare-earth manganite thin films as promising candidate systems for new multifunctional devices.

Interfaces of transition metal oxides are a fertile ground for new physics, often showing novel electronic and magnetic properties that do not exist in the bulk of the materials [1]. Most notably, interfaces of insulating perovskite layers were shown to be conducting [2–4], superconducting [4] and ferromagnetic (FM) [5–7]. Domain boundaries demonstrate equally diverse and fascinating emergent phenomena, and were shown to be conductive [8] or ferroelectric [9]. While the mechanisms of these novel properties are not completely understood, their possible origins include charge transfer, spin-orbital reconstruction, and strain or strain gradient effects [1, 10].

The engineering of multifunctional transition metal oxide interfaces offers vast potential for new discoveries, while simultaneously posing considerable experimental and theoretical challenges [11, 12]. One little-explored direction concerns the interfacial properties of materials such as magnetoelectric multiferroics. Classic examples include orthorhombic (o-) $RE$MnO$_3$ ($RE$ = Tb, Lu, Y), in which a symmetry-breaking magnetic transition causes the direct coupling of antiferromagnetic (AFM) and ferroelectric properties [13, 14]. Surprisingly, studies on both o-TbMnO$_3$ and o-YMnO$_3$ thin films report the existence of a FM magnetization loop [15, 16]. It has been suggested that the ferromagnetism may arise from domain boundaries [17], or alternatively from a uniform canting of an AFM spin structure [18].

To answer the intriguing question regarding the origin of an insulating FM phase in an otherwise AFM system, here we present a study of strained LuMnO$_3$ thin films grown on YAlO$_3$ substrates. The FM magnetization in our films displays a distribution that is peaked at the film-substrate interface. The interface region is where the effects of lattice strain on the crystal properties is largest, and so the lattice strain is identified as the key ingredient for ferromagnetism. We also show the existence of an exchange-field between the ferromagnetism and the long range AFM order also present in the films, which paves the way for studies that are relevant for applications.

Epitaxial thin films of o-LuMnO$_3$ (LMO) were grown on (110) oriented YAlO$_3$ (YAO) [20], with the [1-10] and [001] axes parallel to the substrate surface, and the [110] axis out of plane. Full details of the film synthesis are given in Ref. [21]. By x-ray diffraction the films are shown to be single-phase, untwinned and of single crystalline quality. Rocking scans of the (110) film peaks along the growth direction display a FWHM that is typically $< 0.06°$. For thicker films up to 200 nm the FWHM remains $< 0.10°$. At the out-of-plane (110) film peak, Laue oscillations are observed indicating well aligned film lattice planes. Table I contains film lattice parameters determined by both x-ray and neutron measurements. Compared with bulk o-LMO, on average the film is stretched along the c-axis and compressively strained along the [110] direction [Fig. 1(a)]. This strain pattern is typical for LMO films grown on YAO, and where thinner films show more pronounced strain effects than thicker films [22]. Fig. 1(a) also illustrates the detected monoclinic distortion that arises as a consequence of the substrate-induced lattice distortion [22], and which lowers the crystal symmetry at the interface.

The crystallinity of the LMO thin films was studied using both channelling Rutherford backscattering (c-RBS) and high resolution transmission electron microscopy (HRTEM). C-RBS measurements were done on a $\approx 56$ nm thick film, and using a 2 MeV $^4$He ion beam aligned with the film surface normal in order to study the energy- and hence depth-dependence of the ion channelling [23]. The part of the spectrum due to Lu ions in the sample [Fig. 1 (b)] shows a clear dip in backscattering yield between two regions of higher yield. The dip reflects increased ion channelling, which evidences a coherent and highly crystalline central portion of the film. The two higher yield regions show reduced ion channelling in the vicinity of the film surface (higher channel number) and substrate-film interface (lower channel number), and may be caused by strain, incoherency in the crystal lattice, and crystal defects. By simulating the c-RBS spectrum we identify three main film regions, interface:coherent:surface, with approximate thicknesses 9 nm: 40 nm: 6 nm. However, the boundaries between these regions are likely not sharply defined. Nonetheless, while reduced channelling at the surface is expected due to both random atom distribution and disorder, the reduction close to the substrate-film interface indicates the existence of a broader region of crystal imperfections.

HRTEM measurements were conducted using a FEI´Tecnai F-30 TEM (300 keV) [21]. Fig. 1 (c) shows a HRTEM image of the interface region along the [001] zone axis of a $\approx$ 56 nm thick LMO film. The corresponding diffraction image of Fig. 1 (c) is shown in Fig. 1 (d), and it is found that the positions of the film diffraction spots are located vertically above those of the substrate [inset, Fig. 1 (d)]. These spot locations are different to those expected for bulk o-LMO on the YAlO$_3$ substrate [21]. This shows the illuminated part of the film to contain lattice planes that are both epitaxially-, and compressively-strained to match the underlying substrate lattice along the in-plane [110]-direction. The diffraction image also shows weak intensity haloes; these evidence a degree of film amorphization induced by the focussed ion beam technique used to prepare the films for the HRTEM [21]. Since the amorphization occurs mostly at distances > 10 nm from the direct interface, we focus on the as-grown film crystallinity at shorter distances.

To study the film strain close to the substrate-film interface, images of a strain field contrast are reconstructed by an inverse fast Fourier transformation (FFT) of just the (110) diffraction spots of both film and substrate. The resulting image of the (110) lattice planes [Fig. 1 (e)] shows regions of both continuous and discontinuous contrast at the direct interface, respectively indicating the coherent strain of film lattice planes with respect to the substrate [inset, Fig. 1 (e)], and the presence of local structural incoherence and crystal defects. Figs. 1 (f)- (g) show closer inspections of interfacial regions where disorder is clearly evident. In Fig. 1 (f) the incoherent alignment of the film and substrate lattices result in an extra (110) plane on the film side. In Fig. 1 (g) a misfit dislocation characterized by an extra (110) plane on the substrate side is observed at the direct interface. In these selected regions, dislocations are visible both at the direct interface, and randomly distributed within the explored _10 nm thick volume (marked by ovals). The presence of defects in such a volume is consistent with the reduced ion-channelling observed by the c-RBS measurements. Overall, our study of the film-substrate interface region shows how the lattice mismatch is accommodated at the expense of film crystallinity by both strain and misfit dislocations. Thus, the LMO-YAO interface can be categorized as only semicoherent.

Next we describe the microscopic magnetic properties of the LMO thin films. The antiferromagnetism in a $\approx$ 80 nm thick LMO film was studied using the RITA-II instrument at PSI, Switzerland, and the D10 instrument at ILL, France [21]. The observation of magnetic Bragg peaks [Fig. 2 (a)] is due to the long-range AFM order of the Mn spins in the films. The T -dependence of these peaks indicate the AFM magnetic transition to be _40 K [Fig. 2 (b)]. The magnetic peaks are located at positions described by $\mathbf{G}\pm\mathbf{q}$, where $\mathbf{G}$ is a reciprocal lattice vector, and $q=(0,q_k,0)$ is the magnetic wave vector in reciprocal lattice units. Measurements of the $(0,q_k,1)$ peak evidence incommensurate (IC) magnetic order in the film with $q_k$=0.482(3) [Fig. 2 (a)]. This signals a remarkable change in both the magnetic symmetry, and the AFM order, compared with the commensurate E-type order seen in bulk o-LMO with $q_k$=0.5 [19]. The incommensuration of the magnetism in the film is confirmed by the observation of a weak mirror peak located at (0,0.52,1), and 4 further IC peaks including the $(1,q_k,1)$ peak [Fig. 2 (c)]. A full magnetic structure refinement is not possible from the data, though the intensity distribution of the 6 peaks allows for a feasible bc-cycloid. This proposal is supported by measurements of the $(0,q_k,1)$ peak with the film under $\mu_0$H ∥ a [Fig. 2 (d)], which evidence neither a spin flop transition, nor a change in magnetic symmetry, for fields up to 10 T.

The clearest distinction between the thin film and bulk LMO are differences in the lattice constants, which is indicative of strain that exists throughout the coherent volume of the film. Therefore, our neutron diffraction measurements evidence a strain-induced change of the AFM structure in the films compared with the bulk. To understand this observation, we refer to the magnetic phase diagram reported by Goto et al. [24], where a strong dependence of the AFM order type on the Mn-O-Mn angle is observed. A small increase in the angle would shift the magnetism away from an E-type and towards a cycloidal spin structure, which could happen as a consequence of a strain-induced distortion of the $MnO_6$ octahedra [25]. A similar behavior was observed in $YMnO_3$ where a mixed phase of E-type and cycloidal AFM orders was measured by Wadati et al. [26] and indirectly concluded to be a spiral phase by Fina et al. [27]. It is noteworthy that the differences between the film and bulk crystal structures have a pronounced effect on the magnetic ground state, while the ($T$-dependent) electronic structure remains largely unchanged [25]. The other route to altering the magnetic properties is by changes in the Mn orbital order, and which would support the suggestion that growth-induced strain plays a decisive role in determining the magnetic ground state [25]. Fig. 2(a) shows a further influence of film strain whereby it is seen that the magnetic peak is not resolution limited. The resolution corrected magnetic correlation length is 17(3) nm, and so significantly shorter than the effective film thickness of 66 nm along the direction of $q$.

Next we address both the origin and distribution of ferromagnetism in the LMO thin films. The magnetization profile perpendicular to the film-substrate interface of a $\approx 56$ nm thick LMO film has been determined by polarized neutron reflectometry (PNR) experiments performed at the time-of-flight neutron reflectometer AMOR at PSI. The momentum transfer dependent reflectivity $R(q_z)$ probes the depth-profile of both the chemical composition and the magnetic induction, since both quantities determine the neutron index of refraction. The sensitivity to the in-plane magnetic induction is revealed by differences in the reflectivity curves for neutrons of opposite polarizations (neutron spin parallel (+) and antiparallel (-) to the external field), at the same $q_z$-value.

Fig. 3 (a) shows reflectivity curves for both spin directions taken at $T$=150 K and $T$=10 K, and for $\mu_0$H (001) = 4 T. The dominant contribution to the reflectivity is the structural (i.e. the chemical) composition: the average film density determines the total external reflection up to

$q_z$=0.016 °A$^{-1}$. The interference of the neutrons reflected from both the surface and the LMO/YAO interface leads to the fringes visible at higher qz. The frequency of this oscillation gives the expected film thickness of t=56 nm. At 150 K, $R(q_z)$ for both spin directions are equal. On cooling, a difference between spin up and down is first detected at ≈ 100 K, and is largest at the lowest measured $T$ of 10 K. For clarity, in Fig. 3 (b) we present the normalised ratio (R$^+$-R$^-$)/0.5(R$^+$+R$^-$) for the 10 K measurement.

To analyse the PNR measurements quantitatively, the internal magnetization distribution was modelled by the following gaussian profile: $\mu_{Mn}(z)$=1.07$\mu_B$ exp[-(z/39 nm)$^2$], with the peak at z = 0 located at the interface. Using this profile, Fig. 3 (b) shows the excellent agreement between the simulation (red line) and the experimental data. In comparison, the simulation with the same gaussian function centred at the film surface (black line) is in clear contradiction to the data. According to the gaussian profile, 50% of the integrated magnetisation exists within the first 18 nm of the film, and the average moment per Mn of <$\mu_{Mn}$>≈0.50(5)$\mu_B$. Other model functions, such as a gauss error function, were tried for simulating the data. While the precise details of the moment decay are model dependent, the essential details of the magnetization distribution were captured by all model functions; a FM interface layer typically ≈10 nm thick with $\mu_{Mn}$≈1.1$\mu_B$, and the moment decaying towards the film surface.

Ferromagnetism in the films is also confirmed by standard bulk measurements of the $M(H)$ hysteresis loops at 5 K [Fig. 3 (c)]. The zero field-cooled $M(H)$ loop reaches a maximum by 3 T [21], and gives an average magnetization across the film of <$\mu_{Mn}$>=0.485(5)$\mu_B$ which agrees well with that estimated from PNR. Furthermore, by comparing between the zero field-cooled $M(H)$ loop (black curve) and a loop measured after field-cooling in 4 T (red curve), a significant shift of 205 Oe in the field cooled loop is noted, and signals a clear exchange-bias between the FM and AFM fractions in the sample. The exchange field therefore establishes the existence of coupling between the two magnetic phases in our LMO thin films, which is a crucial property for applications.

In Fig. 3 (d), we summarize schematically the expected magnetic situation in the LMO thin films. The FM magnetization displays e.g. a gaussian profile that is peaked at the interface, and which is much reduced at the film surface. On moving away from the interface region, the magnetism evolves towards long-range, likely cycloidal, incommensurate AFM order. Since the precise deviation of the AFM order away from the E-type structure is sensitive to the strain in the films, it may be expected that the E-type would be present, or even dominant, with increasingly thicker films. Consequently, a single sample may simultaneously host different magnetic structures that could be distinguished in one experiment. The illustration of the single phase material shown in Fig. 3 (d) is equivalent to an artificial structure described in Ref. [28] and realized experimentally in Ref. [29].

A FM magnetization distribution that is peaked at the LMO/YAO interface strongly excludes that it originates from within crystal domain boundaries. This is because few domain boundaries are expected in our twin-free films, which is quite unlike $RE$MnO$_3$ thin films grown on (100) SrTiO$_3$, and in which a uniform magnetization distribution across the film was identified using PNR [30]. Instead, the FM magnetization in the LMO films is peaked where the strain gradient is largest, and where the film hosts regions of epitaxially-strained lattice planes with monoclinic distortion [22], and strain induced crystal defects. On moving away from the interface, the decay of the FM magnetization is concomitant with the shallowing lattice strain gradient. All of these observations strongly tie the origin of the FM magnetization to strain-induced alterations of the crystal properties which are most pronounced close to the interface. Nevertheless, the microscopic origin of the strain-induced ferromagnetism remains to be addressed. One possibility is that the monoclinic distortion of the orthorhombic lattice, which is also largest at the interface [22], could provide the necessary reduction of crystal symmetry that alters the orbital ordering and allows a uniform magnetization to be manifested independently of the bulk AFM order. Alternatively, the strain-induced misfit dislocations may locally host small variations in stoichiometry that display FM properties. These proposals can be tested by XMCD experiments.

Our experimental results could also be explained by a strain-induced change in the relative strength of the exchange interactions. However, since none of the bulk materials show macroscopic moment FM phases, and the exchange along the c-axis is solidly AFM, we think that such a scenario is unlikely. Further, this proposal would also not explain why similar FM properties have been observed for a number of different $RE$MnO$_3$ films grown on a range of different substrates [18, 30]. An alternative possibility is that the growth-induced strain causes a canting of the AFM phase below TN and hence a projection of the FM moment is observed. This could explain the reduced Mn moment of $\approx$1$\mu_B$ at the interface. However, the FM moment is first detected by PNR well above TN at $\approx$100 K. Moreover, the measured exchange field does not support spin canting as the origin for the observed FM signal.

To summarize, we have shown unequivocally that LuMnO3 thin films grown on YAlO$_3$ substrates display a large moment ($\approx$1$\mu$B) ferromagnetic magnetization that is peaked at the substrate-film interface. The origin of the ferromagnetism is shown to be tied to strain-induced alterations of the crystal properties which are most pronounced close to the interface. We also demonstrate the directly-coupling of the ferromagnetism with the incommensurate antiferromagnetic order also present in the same phase below $\approx$40 K. Our approach comprises a novel route for generating coupled magnetic orders in a single phase thin film. Such effects should also be present at room temperature and our study paves the way for the development of new multi-function devices.

Work at the PSI is supported through both the SNF, and the NCCR program MaNEP. Neutron experiments were performed at the Swiss Spallation Neutron Source, SINQ, PSI, Switzerland and at the ILL in Grenoble, France. We also acknowledge support by the Electron Microscopy Center of ETH Zürich (EMEZ). We thank Ch. Bernhard, M. Fiebig, J. Fontcuberta, and M. Laver for valuable discussions.

*christof.schneider@psi.ch

TABLE I. The lattice constants of both the YAlO₃ substrate, and the LuMnO₃ (LMO) films determined by both x-rays and neutrons. Values for bulk LMO are taken from Ref. [22]. All values are given for T =298 K, and the *Pbnm* space group.

| | YAlO₃ substate | LMO (x − rays) | LMO (neutrons) | LMO (bulk) |
|---|---|---|---|---|
| a (Å) | 5.18 | 5.21 | 5.19 | 5.20 |
| b (Å) | 5.31 | 5.73 | 5.76 | 5.79 |
| c (Å) | 7.35 | 7.36 | 7.34 | 7.30 |

**Figure caption**

**Fig. 1:** (a) Sketch of the substrate-induced rhomboidal distortion of the ab-plane in thin film LMO, as compared with the undistorted lattice of bulk o-LMO. (b) The c-RBS spectrum of the LMO thin film on the YAO substrate, focussing on the contribution due to Lu. Black dots represent the raw data. The red triangles are the simulation. (c) A HRTEM image of a LMO thin film on the YAO substrate at 390 kx magnification. (d) The corresponding diffraction image of panel (c). (e) The reconstructed image of the (110) lattice planes after an inverse FFT. The inset blue window highlights a region of coherently strained film-substrate lattice planes. (f) and (g) show magnified images of the green and red regions in panel (e). Ovals highlight crystal defects.

**Fig. 2:** (a) The (0 $q_k$ 1) magnetic peak in a LuMnO₃ thin film. A gaussian fit of the main peak gives the peak centre at $q$=(0 0.482 1). The solid bar in (a) represents the instrumental resolution of RITA-II. (b) The $T$ –dependence of the $q$=(0 0.482 1) peak, with lines as guides for the eye. (c) The q=(1 $q_k$ 1) peak measured at D10. (d) The $\mu_0H$-dependence of the

$q$=(0 0.482 1) peak for $\mu_0$H $\parallel$ a up to 10 T. Empty (filled) symbols denote data taken T=2 K (T=50 K). Curves are displaced vertically for clarity, and peaks are fitted with gaussian line-shapes.

**Fig. 3: a)** (a) Intensities from PNR measurements as a function of $q_z$ at 10 K and 150 K. The latter is multiplied by 0.1 for clarity. (b) The normalized spin up/down ratio at 10 K, and fitted by a gaussian profile for the magnetization that is peaked at the substrate-film interface (red line) and at the surface (black line). (c) $M(H)$ loops for the sample measured after zero-field cooling to 5 K (black), and after field-cooling to 5 K in 4 T (red). (d) Sketch of the established magnetic situation in the LMO film. The FM layer located near the strained film-substrate interface evolves towards a likely cycloidal AFM order when moving down the strain-gradient towards the film surface.

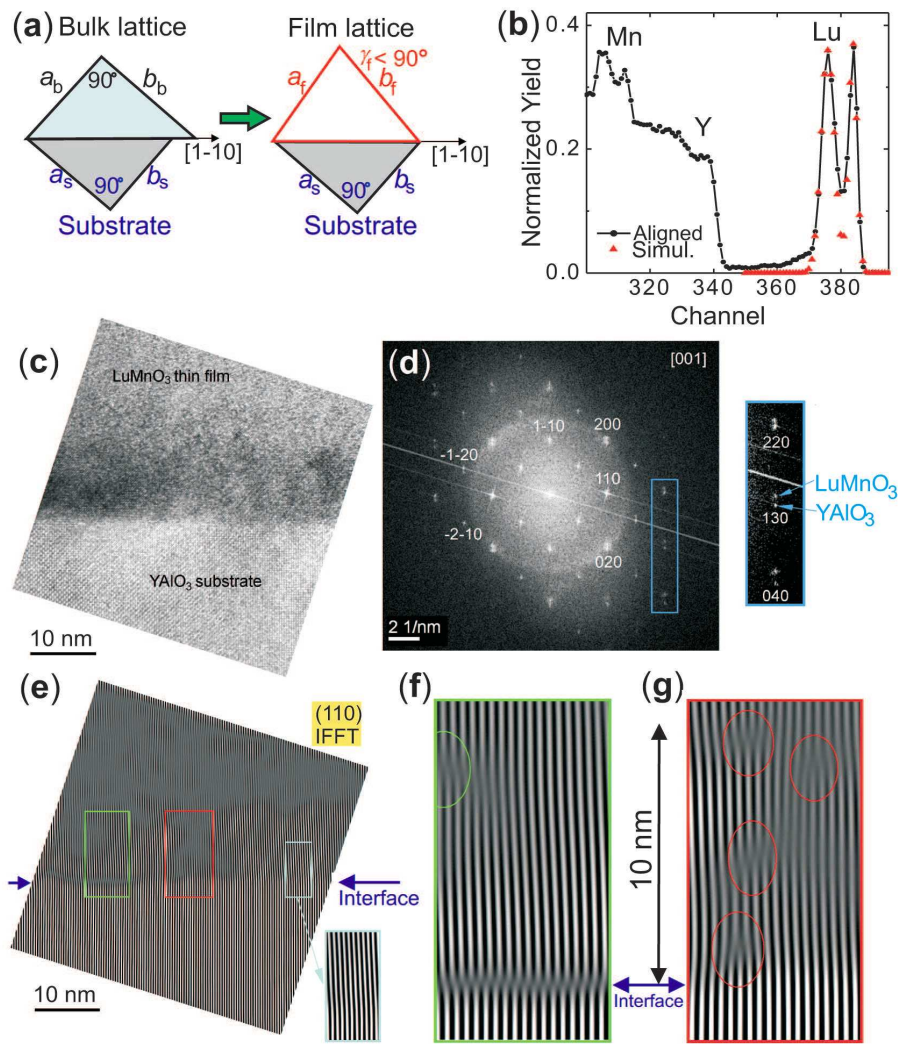

Fig. 1: White et al.

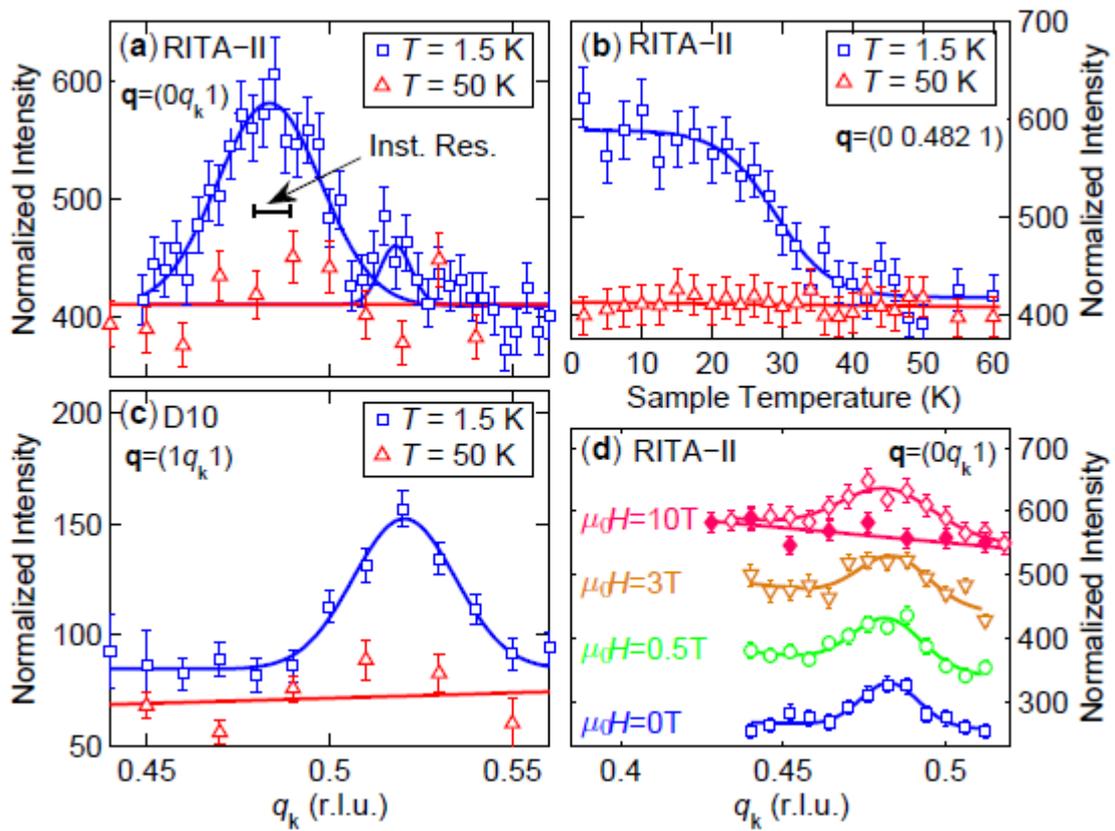

Fig. 2: White et al.

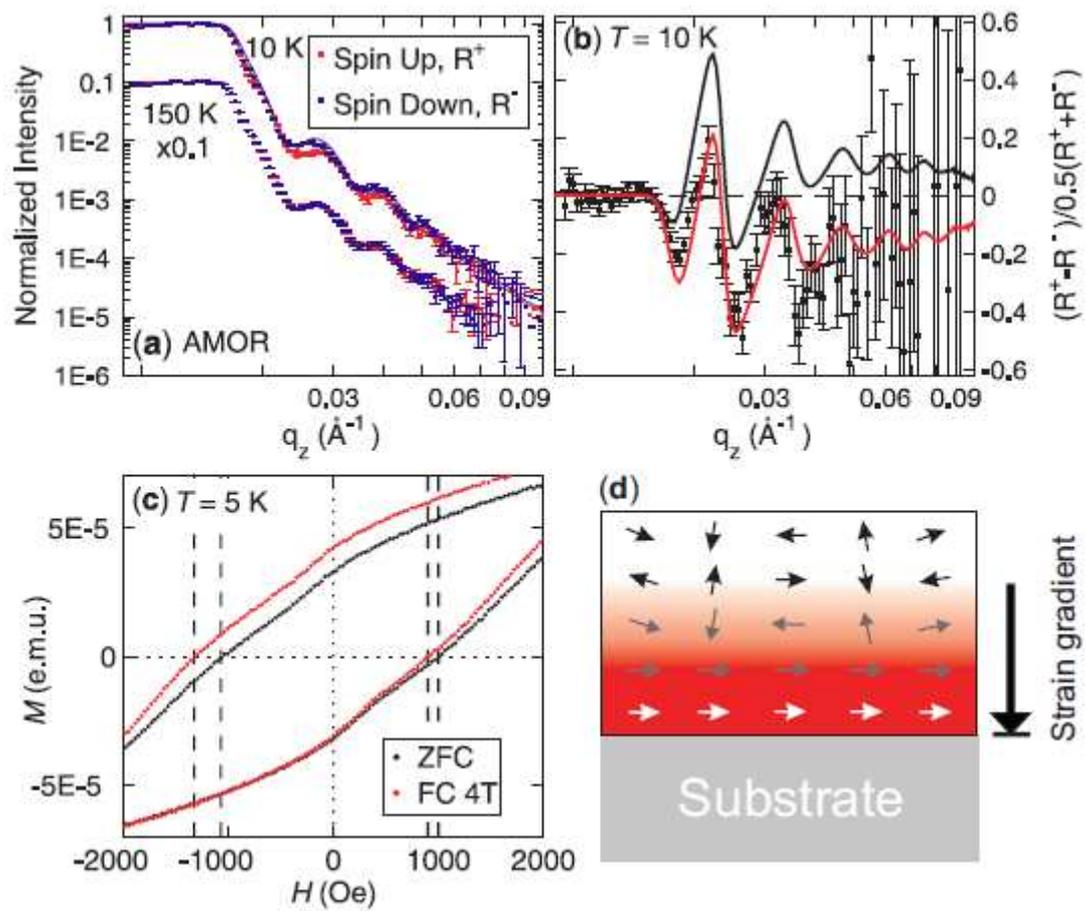

Fig. 3: White et al.

Supplementary Material to "Strain-induced ferromagnetism in antiferromagnetic LuMnO₃ thin films"

J.S. White,[1,2] M. Bator,[1] Y. Hu,[1] H. Luetkens,[1] J. Stahn,[1] S. Capelli,[3] S. Das,[4] M. Döbeli,[5] Th. Lippert,[1] V.K. Malik,[4] J. Martynczuk,[6] A. Wokaun,[1] M. Kenzelmann,[1] Ch. Niedermayer,[1] and C.W. Schneider[1,*]

[1] Paul Scherrer Institut, CH-5232 Villigen, Switzerland
[2] Laboratory for Quantum Magnetism, Ecole Polytechnique Fédérale de Lausanne (EPFL), CH-1015 Lausanne, Switzerland.
[3] Institut Laue-Langevin, 6 rue Jules Horowitz, 38042 Grenoble, France.
[4] University of Fribourg, Department of Physics and Fribourg Centre for Nanomaterials, Chemin du Musée 3, CH-1700 Fribourg, Switzerland
[5] Labor für Ionenstrahlphysik, Eidgenössische Technische Hochschule Zürich, CH-8093 Zürich, Switzerland
[6] Electron Microscopy Center of ETH Zurich (EMEZ), Eidgenössische Technische Hochschule Zürich, CH-8093 Zürich, Switzerland

## EXPERIMENTAL DETAILS

Epitaxial thin films of LuMnO₃ (LMO) were grown on (110) and (100)-oriented YAlO₃ (YAO) single crystalline substrates [1] at $T_S = 760°C$, and by pulsed laser deposition(PLD) using a KrF excimer laser ($\lambda = 248$ nm, 2 Hz) with a laser fluence of F = 3 J/cm². To provide more atomic oxygen for the film growth, either N₂O was used as background gas ($p$N₂O=0.2 mbar) or the laser pulse was synchronized with a gas pulse to inject N₂O into a laser induced O₂ plasma (O₂ background: $p$=0.2 mbar) [2]. For preparing the thin films, stoichiometric sintered ceramic targets of hexagonal LMO were used, and which were prepared by conventional solid-state synthesis. For the experiments reported in this study, the film thicknesses ranged from ≈56 - 80 nm.

The high resolution transmission electron microscopy (HRTEM) measurements were conducted using a FEI Tecnai F-30 TEM. The film was prepared for the measurements by the focused ion beam (FIB) technique on a CrossBeam NVision 40 from Carl Zeiss with a gallium liquid metal ion source, a gas injection system and a micromanipulator MM3A from Kleindiek. After electron beam deposition of carbon in SEM mode at low scan speeds, thin films were protected by a carbon cap and lamellae were cut free with trenches from both sides with 13 and 3 nA at 30 kV. After the lift-out was performed the lamellae were polished to ion transparency with currents down to 10 pA at 30 kV. The amorphization was diminished by low kV showering for several seconds at 5 and 2 kV.

The neutron diffraction experiments were performed using the triple-axis spectrometer RITA-II at the PSI in Villigen, Switzerland and the D10 four-axis diffractometer at the ILL in Grenoble, France. In these experiments a 80 nm thick LMO film was investigated. At RITA-II, the LMO film was mounted in the ($0k0$)-($00l$) scattering plane, and an incoming neutron energy of 4.6 meV was chosen. A PG filter between monochromator and sample, and a cooled Be filter between sample and analyzer were installed to suppress higher order contamination.

Polarized neutron reflectometry probes the depth profile of both the chemical composition and of the magnetic induction. Both quantities determine the neutron index of refraction. A grazing incident beam is refracted and partially reflected at the interfaces. The interference of all (multiply) reflected beams leads to a momentum transfer dependent reflectivity $R(q_z)$. The experimental data were analyzed by comparing the results to numerical simulations. PNR was performed on a $\approx$56 nm LMO film at the time-of-flight neutron reflectometer AMOR at PSI.

The magnetization results were obtained using a conventional PPMS from Quantum Design. Measurements were conducted after either a zero-field- or field-cooling to 5 K, and on subsequently sweeping the magnetic field. Fig. 1 shows the $M(H)$ loop measured after zero field cooling, where it is seen that the maximum value of 0.485 $\mu_B$ is obtained under an applied field of 3 T. Polarized neutron reflectometry shows that the maximum magnetization per Mn atom is $\approx$1$\mu_B$. This magnetization extends mainly over the first 10 nm into the film away from the film/substrate interface and, according to the profile given in the main text, yields an integrated value of $\approx$0.5 $\mu_B$ per Mn atom.

**EPITAXIALLY STRAINED FILM LATTICE**

In the main paper, we presented a high-resolution transmission electron microscopy image obtained from close to the substrate-film interface, and along the (001) zone axis [Fig. 1(c) in the main paper]. The corresponding diffraction image shown in Fig. 2(a) (same as Fig. 1(d) in the main paper), and careful examination shows the film diffraction spots to be located vertically above the substrate spots. These spot locations lie in contrast to the relative positions expected for bulk o-LMO and the substrate [Fig. 2(b)]. Therefore, the alignment of the film diffraction spots along the [1-10]-direction (the out of plane direction of the film) signifies that the illuminated part of the film lattice is epitaxially strained to match the underlying substrate lattice along the in-plane [110]-direction.

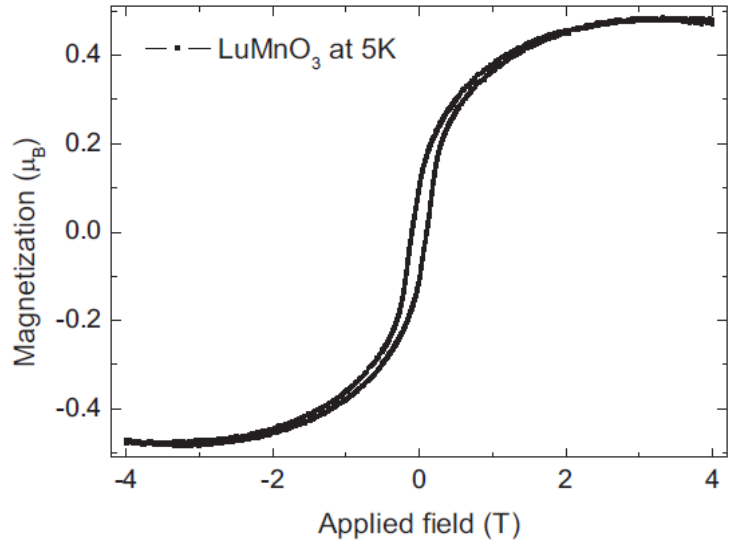

FIG. 1. Zero field cooled $M(H)$ loop measured at 5 K on the LMO thin film.

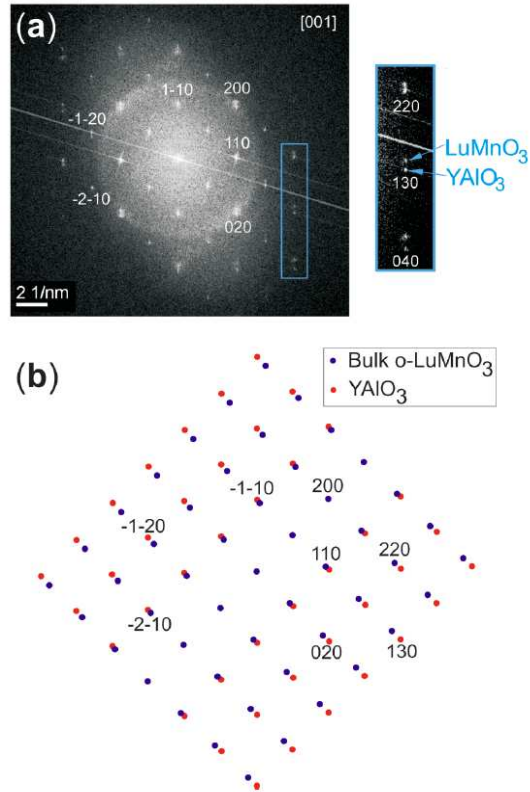

FIG. 2. (a) The diffraction image obtained from the high resolution transmission electron microscopy image shown in Fig. 1(c) of the main paper. It shows the film diffraction spots to be located vertically above the substrate diffraction spots. This is shown in more detail by the inset. The location of the film spots is in contrast to that expected for the orthorhombic lattice LMO shown in panel (b). This shows that the studied part of the film is epitaxially strained to match the substrate lattice.